**Hybrid confinement techniques for polariton simulators**

Johannes Düreth [a*§], Philipp Gagel [a§], David Laibacher [a], Oleg A. Egorov [b], Simon Widmann [a], Simon Betzold [a], Monika Emmerling [a], Siddhartha Dam [a], Alexia Landry [a#], Christian G. Mayer [a], Martin Kamp [a], Aniela Woyciechowska [c], Barbara Piętka [c], Ulf Peschel [b], Sven Höfling [a], Sebastian Klembt [a*]

[a] Lehrstuhl für Technische Physik, Wilhelm-Conrad-Röntgen-Research Center for Complex Material Systems, and Würzburg-Dresden Cluster of Excellence ct.qmat, Julius-Maximilians-Universität Würzburg, 97074 Würzburg, Germany

[b] Institute of Condensed Matter Theory and Optics, Friedrich-Schiller-Universität Jena, Max-Wien Platz 1, 07743, Jena, Germany

[c] Institute of Experimental Physics, Faculty of Physics, University of Warsaw, ul. Pasteura 5, 02-093 Warsaw, Poland

[*] E-Mail: johannes.duereth@uni-wuerzburg.de,
[*] E-Mail: sebastian.klembt@uni-wuerzburg.de

[§] J.D. and P.G. contributed equally to this work

Keywords: Exciton-polaritons, photonic lattices, quantum simulation, polariton condensation, microcavity lattice design

*Exciton-polariton III-V semiconductor microcavities provide a robust platform for emulating complex Hamiltonians, enabling topological photonics and quantum simulation for advanced photonic functionalities. Here, we introduce two novel fabrication techniques – etch-and-oversputter and deposit-and-oversputter – that overcome limitations of traditional photonic confinement. Both use structured, locally elongated semiconductor cavities to create deep, highly controllable potentials, while leveraging high-quality GaAs-based materials, which achieve excellent Q-factors. A sputtered all-dielectric top mirror introduces an innovative hybrid approach, simplifying fabrication while maintaining quality compared to deep ion etching. Utilizing a Kagome lattice as a benchmark, we show high-quality optical band structures previously inaccessible with deep etching. Furthermore, we study a two-dimensional breathing Kagome lattice and demonstrate polariton lasing from a zero-dimensional corner mode, confirming precise control over couplings and tight polariton localization. These methods enable fabrication of intricate lattices, including higher-order topological insulators, or on-chip quantum regimes utilizing the polariton blockade mechanism due to tight photonic confinement.*

## 1. Introduction

In recent years, two-dimensional polariton lattices,[1–3] have emerged as novel and powerful platforms for the emulation of complex Hamiltonians,[4–6] topological photonics,[7–9] while providing a technologically viable access to quantum light.[10–13] These lattice simulators enable the observation of properties such as intensity, phase, energy, angle and coherence of the evolving wave packets directly using optical methods. Microcavity exciton-polaritons (polaritons),[14,15] have raised particular interest because of their low effective mass and composite bosonic nature with large optical non-linear properties. Along this line, dynamic Ising machines,[5] and dynamical Floquet lattices have been realized,[16] polariton quasicrystals[17] and potential for neuromorphic computing are being explored.[18–20]

Established photonic trapping methods, such as etched micropillars,[2,7,8] face significant limitations when applied to densely packed lattices, such as square or hexagonal lattices. Achieving homogeneous and large etch depth, and concurrently small-sized features within a micropillar lattice is an intricate challenge. Consequently, the creation of homogenous trapping potentials with existing methods is severely constrained. To address this, up to now the etch-and-overgrowth (EnO) technique[21] proved to be a powerful tool for the creation of large potential landscapes[22–24] consisting of coupled resonators. A similar technique has been successfully applied to dielectric cavities filled with perovskites. While this technique allows the creation of well-defined polariton lattices at room temperature,[9,25–27] the overall material quality, in particular the intrinsic exciton linewidth and birefringence, restrict perovskite systems from reaching the performance of GaAs microcavities, especially for demanding applications such as quantum-light generation from lattices.[28]

For the fabrication of highest-quality GaAs-based lattices, the EnO method involves interrupting the growth process after the cavity layer, etching the photonic traps directly into the cavity layer and then continuing the growth process using the same materials. This exploits the fact that a thinner cavity acts as a potential barrier for the cavity photons. Typical confinement potentials exceeding $E_{\text{pot}} = -10$ meV can be achieved with an etch depth of $\sim 10$ nm, offering significantly improved homogeneity compared to the deep micropillar etching, requiring depths of several micrometers. In addition, outstanding control of waveguide,[29] or circular trap coupling[22–24] have been achieved by precisely tailoring feature distance and etch depth of the cavity. However, a key limitation of the EnO technique is its low flexibility. The layout is typically etched across a full wafer, since fitting individual pieces to match the natural radial thickness gradient in molecular beam epitaxy (MBE) is hard to achieve. To overcome these challenges, we propose to replace the crystalline top distributed Bragg reflector (DBR) with a sputtered dielectric mirror,[30–32] for both the etch-and-oversputter (EnS) and deposit-and-oversputter (DnS) technique. Additionally, for the DnS method, the etching step is replaced by the sputter deposition of $TiO_2$, which induces confinement through local elongation of the cavity.[33] Together, these developments define a new hybrid fabrication technique that substantially increases flexibility while preserving the homogeneity of the photonic potential landscape observed for crystalline EnO structures. The EnS and DnS approaches thus represent major advancements in the fabrication of complex photonic potentials, opening new opportunities for the development of advanced topological devices.

Firstly, we compare the polariton band structures of Kagome lattices using three different fabrication methods: the established etch-and-overgrowth method, the novel etch-and-oversputter and deposit-and-oversputter approach. This comparison is performed using momentum- and real-space spectroscopy. Subsequently, we analyze the confinement potentials through the solutions of the two-dimensional Schrödinger equation. Finally, we demonstrate the versatility of the EnS method by implementing a complex breathing Kagome lattice hosting corner modes, and evidence polariton lasing from these modes.

## 2. Results and Discussion

The experimental results presented in **Figure 1** are obtained using the two new fabrication techniques: etch-and-oversputter (cf. **Figure 1e**-**1h**), and deposition-and-oversputter (cf. **Figure 1i**-**1m**). Both methods are similar to the established etch-and-overgrowth (cf. **Figure 1a**-**1d**) technique as all three methods locally alter the cavity length to create the confinement potential.[21–24]

For clarity, we detail the fabrication and working principles of the EnS technique using the specific structure shown in **Figure 1e** and **1f**, which provide an overview and a schematic of the layer architecture. Fabrication starts with an epitaxial bottom DBR, consisting of 30 $Al_{0.15}Ga_{0.85}As$/AlAs mirror pairs and a $\lambda$-cavity. At the design wavelength, the electric field intensity in the $\lambda$-cavity exhibits maxima at its center and at the interfaces to the top/bottom DBR. Therefore, two stacks of three 13 nm wide GaAs quantum wells embedded in 10 nm $Al_{0.30}Ga_{0.70}As$ barriers are positioned at the cavity center and the interface to the lower DBR. After epitaxial growth of these layers, the lattice potential landscape is defined using electron beam lithography and wet etching into the cavity layer to create micron-sized mesas (cf. SI Section S5). Finally, a top DBR consisting of 12 pairs of $SiO_2$/$TiO_2$ is sputtered. The local variation of the cavity length of a few tens of nanometers is imprinted into this top DBR. An electron microscopy image of the EnS sample is provided in SI Section S2.

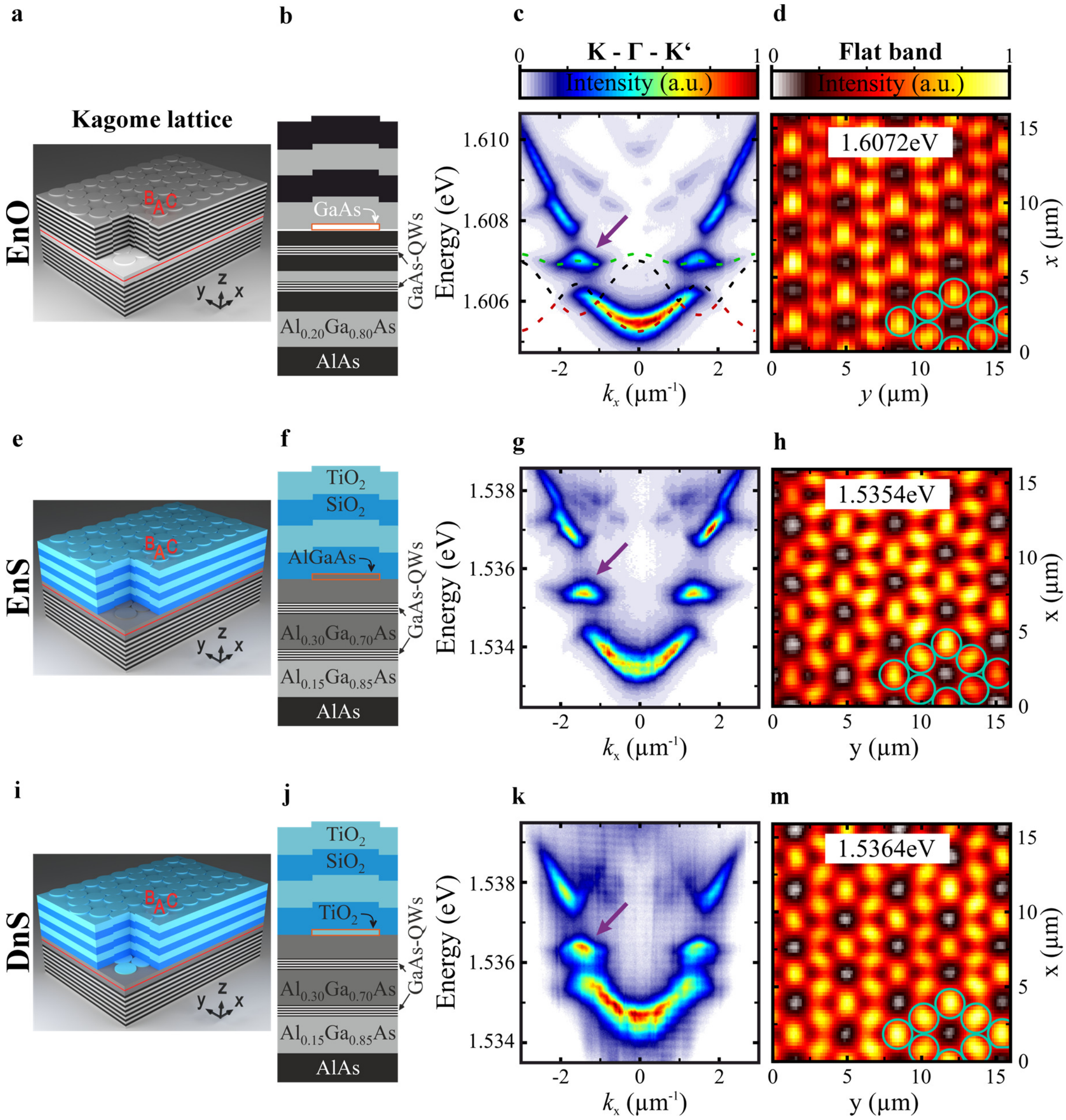

*Figure 1*: ***Comparison of etch-and-oversputter (EnS) and deposit-and-oversputter (DnS) methods to the established etch-and-overgrowth (EnO) method. a****, **e** and **i** Artistic impressions of the Kagome lattice for the EnO, EnS and DnS technique, respectively. The characters A, B and C denote sites on a different sublattice. A sketch of the epitaxial layout for the EnO, EnS and DnS method is shown in **b**, **f** and **j**, respectively. The main difference is the choice of top mirror and fabrication method of the mesas. The epitaxial top mirror in the case of EnO is substituted for a dielectric mirror for EnS/DnS (light blues). The mesa is highlighted in an orange box. **c**, **g** and **k** Momentum-space spectra of the Kagome band structure along the high symmetry direction K-Γ-K' for the **c** EnO, **g** EnS and **k** DnS sample. The dashed lines show a tight-binding model while violet arrows highlight the flatband. **d**, **h** and **m** Real space mode tomography of the flatband of the Kagome lattice in a **d** EnO, **h** EnS and **m** DnS sample. The cyan circles represent the underlying lattice potential.*

The DnS technique eliminates the etch step in the fabrication process entirely. Instead, the mesas are defined through sputter deposition of $TiO_2$ onto a mask (see **Figure 1i-1j**). This approach has the benefits of leaving the epitaxial layer completely untouched, avoiding etching defects and reducing the surface roughness in comparison to the EnS method. A comparison between the EnO and EnS/DnS layer structures, shown in **Figure 1b** and **1f/1j**, respectively, highlights the main difference: the material used for the top DBR. The dielectric DBR offers several advantages, such as a wider stopband while preserving the Q-factor. Furthermore, monocrystalline overgrowth is not needed, lifting material restrictions and eliminating elaborate post-etch cleaning, greatly reducing fabrication complexity. The minimal radial thickness gradient of the sputtered layers and the wider stopband enable processing of individual wafer pieces, substantially reducing iteration times. Importantly, the cavity thickness gradient that controls mode detuning remains unchanged since the cavity material is grown using MBE. A basic characterization of the EnS and DnS samples, as well as the composition of the EnO sample is presented in the SI Sections 2-4.

Compared to conventional confinement methods like deep etching, a notable advantage of the EnO technique is its ability to create homogeneous potential landscapes even for complex lattice geometries.[34] To evaluate this, we compare the band structure and local mode distribution achieved with the EnO, EnS and DnS techniques, using the Kagome lattice as a representative example. In each case, the local elongation of the cavity layer – $Al_{0.30}Ga_{0.70}As$ for EnS; GaAs for EnO; and $TiO_2$ for DnS, – key to this confinement technique, is visible. The Kagome lattice is constructed from circular mesas with a diameter of $d = 2$ µm, center-to-center distance $a = 2$ µm and a reduced distance of $v = a/d = 1$, indicating touching mesas.

To assess the quality of the band structures, momentum-space measurements (cf. SI Section S1) along the K-Γ-K' direction are presented in **Figure 1c**, **1g** and **1k**. Additionally, **Figure 1c** shows exemplary tight-binding calculations including next-nearest neighbor hopping. All momentum-space spectra clearly resolve the *s*-bands as well as parts of the *p*-bands with high quality. The second *s*-band is not visible in the first Brillouin-zone due to destructive interference.[2] The Dirac-cone is located at the $K/K'$-points, where the second and first *s*-band intersect. The characteristic flatband is located above the first and second *s*-band and marked by a violet arrow. For the ground state at$k = 0$ µm$^{-1}$, the linewidth of the DnS sample is measured to be $\delta E_{\mathrm{DnS}} = 680$ µeV, slightly larger than the $\delta E_{\mathrm{EnS}} = 500$ µeV, which is comparable to the linewidth of $\delta E_{\mathrm{EnO}} \approx 400$ µeV. However, we attribute these minor differences to variations in epitaxial growth and the positioning of the stopband in the sputtering process. **Figure 1d**, **1h** and **1m** show the flatband mode distribution obtained via real-space scanning. In all cases, the flatband clearly reveals the underlying lattice potential (cyan circles) and demonstrates high spatial homogeneity. The slightly lower resolution observed for the EnO sample is due to the larger step size used during the photoluminescence scan.

Overall, the results in **Figure 1** show that the EnS and DnS techniques can readily compete with established methods in the quality of the photonic band structure and homogeneity of the trapping potential, even in the case of complex potentials like the Kagome lattice. Additionally, the EnS and DnS techniques offer the advantage of shorter iteration cycles, providing greater flexibility for optimizing parameters in new lattice geometries.

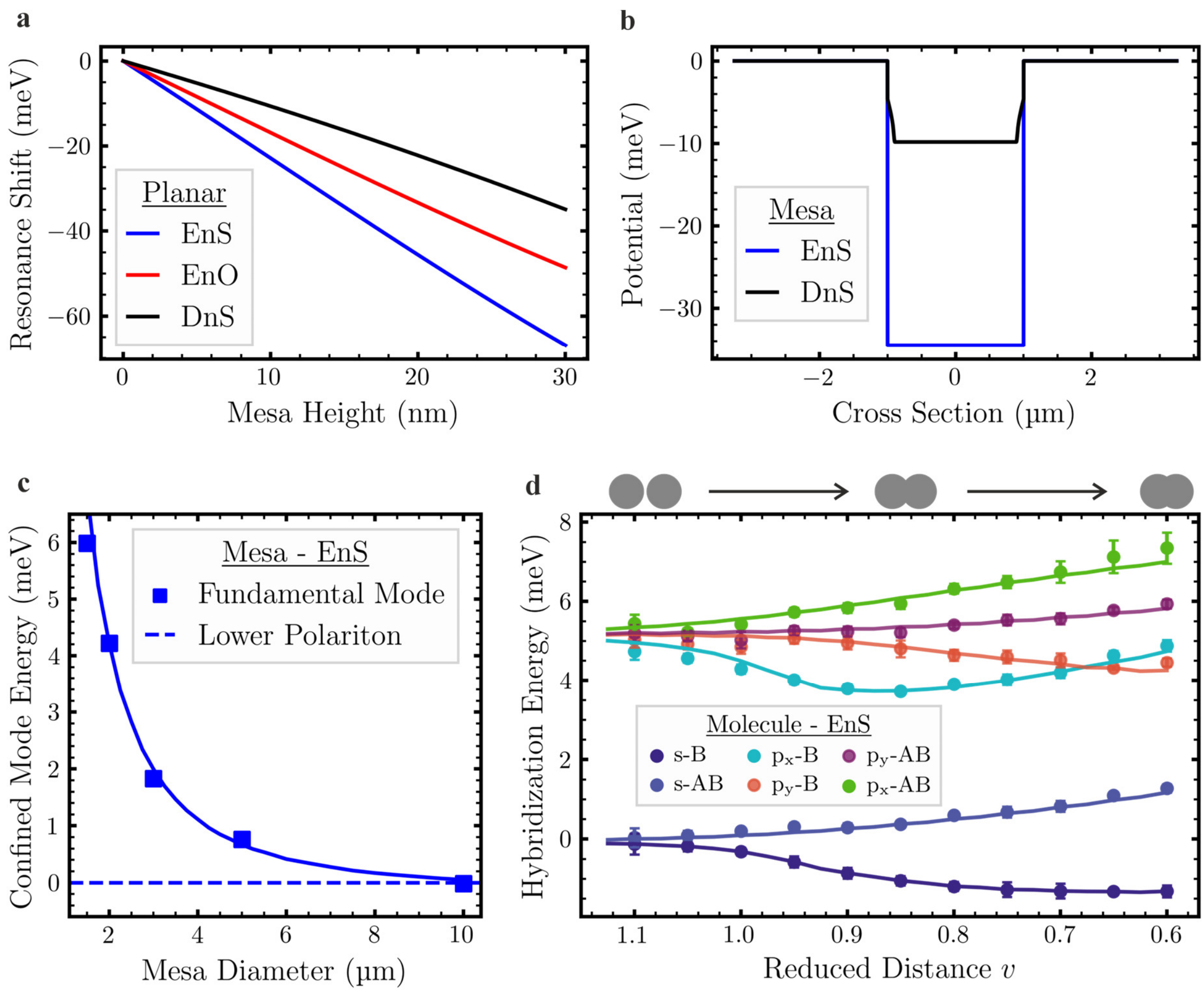

*Figure 2:* ***Theoretical and experimental analysis of the properties of the potentials in the EnS and DnS method highlighting the experimental control over the technology.*** ***a*** *shows transfer matrix calculations of the resonance shift of a quasi-planar microcavity in the three different samples. The magnitude of the resonance shift is equivalent to the potential depth shown in* ***b*** *for a particular Mesa height. Panel* ***c*** *shows the experimentally measured fundamental mode energy of mesas with different diameters (boxes), as well as the simulated eigenmode energy. In* ***d*** *the spectrum of two* $d = 2\ \mu m$ *wide proximity coupled mesas is shown, including s-type and p-type modes. Solid lines correspond to theoretical simulations based on a 2D Schrödinger equation. The EnS sample is different from the one studied in Figure 1 and Figure 3, with an etch depth of* $h_{EnS} = 15.3$ *nm.*

We now examine the mechanism of photonic confinement in greater detail. In all cases, a relative local elongation, induced by etching or material deposition, leads to a red shift of the cavity mode relative to the shorter regions, thereby creating the photonic potential (cf. SI S2-S4). This elongation is visible in the cross-sectional sketches of an individual lattice site in **Figure 1b**, **1f** and **1j**. Furthermore, the spatial extent of the unetched or deposited regions imposes additional, lateral confinement. This manifests in a blueshift of the photonic eigenmodes of these mesa structures. In the EnS sample, a $h_{\mathrm{EnS}} = 13.4$ nm etch-depth results in an experimentally determined photonic potential depth – defined as the energy difference between photonic modes inside and outside of large mesas with a diameter $d > 30$ µm – of $E_{\mathrm{trap}}^{\mathrm{EnS}} = 22.4$ meV. In contrast, the DnS method produces a comparatively smaller potential depth of $E_{\mathrm{trap}}^{\mathrm{DnS}} = 9.6$ meV for a $h_{\mathrm{DnS}} = 13.5$ nm deposited $TiO_2$ layer. This difference is theoretically verified by transfer matrix method calculations for planar cavities for all three fabrication methods. **Figure 2a** presents simulation results of the photonic resonance shift in a planar cavity (neglecting absorption) which corresponds to the difference in energies off- and on a large mesa. It is evident that the photonic mode energy shifts differently for all three techniques, even when accounting for the optical path length difference. This resonance shift is proportional to the maximum potential depth of a mesa.

The spectral resonance of a microcavity is governed by the condition for constructive interference, which requires the phase accumulation for a round-trip within the cavity to be a multiple of $2\pi$. For a planar cavity under normal incident this yields $2k_{\mathrm{in}}L_{\mathrm{cav}} + \varphi_1 + \varphi_2 = q2\pi$, where $k_{\mathrm{in}} = n_{\mathrm{in}}2\,\pi/\lambda_0$ is the incident wavevector, $n_{\mathrm{in}}$ the refractive index of the incidence medium, $\lambda_0$ the photonic resonance wavelength in vacuum, $L_{\mathrm{cav}}$ the physical cavity length, $\varphi_1$ and $\varphi_2$ the phases of the reflection amplitudes of the lower and upper mirror, respectively.[35]

The phases of the reflection amplitudes depend on the type of DBR and the refractive indices of the constituent materials, as well as wavelength of the incident light. On the III-V semiconductor platform, the lower refractive index contrast of the DBR increases the reflection phase and leads to a redshift of the cavity resonance wavelength at a given cavity length when compared to EnS. Therefore, the slope of the resonance shift of EnS (blue line) in **Figure 2a** is larger than the slope for EnO (red line). Moreover, the dependence of the reflection phase on the refractive index of the cavity material varies with DBR type. For L-type DBRs (start with low-index material), the reflection phase decreases with increasing refractive index of the cavity material as $\propto 1/n_{\mathrm{in}}$.[35]As a result, for a given change in the optical path length, the spectral shift of the cavity is smaller for DnS compared to EnS, due to the deposited $TiO_2$ (both are L-type DBRs). Conversely, for H-type DBRs (start with high-index material) the reflection phase increases linearly with $n_{\mathrm{in}}$ .[35]

When the lateral mesa dimension is reduced, the eigenmodes of the microcavity transition from a quasi-continuum to distinct, localized modes. These potentials are well approximated by finite rectangular wells in 1D. The EnS sample studied has an etch depth of $h_{\mathrm{EnS}} = 15.3$ nm. **Figure 2b** illustrates a potential profile for a mesa of diameter $d = 2$ µm, demonstrating the depth and shape difference between the EnS and DnS methods. The DnS pillar shape is modelled according to J. Arscott, [36] taking angled deposition and shadowing effects into account.

The mesa eigenenergy can be tuned over one order of magnitude by varying their size, as shown in **Figure 2c**. The blue squares indicate the measured energy of the fundamental eigenmode in the mesa relative to the energy of the lower polariton without additional confinement in the *xy*-plane (dashed blue line). At a mesa size of $d = 2$ µm, the confined eigenmode exhibits a blueshift of approximately $\Delta E = 4.25$ meV. This tunability enables the emulation of Hamiltonians that require different on-site energies, such as the quantum valley Hall effect in an unbalanced honeycomb lattice, emulating e.g. hexagonal boron nitride or transition metal dichalcogenides.[27,37]

Another key parameter for Hamiltonian emulation is the hybridization energy of two proximity-coupled mesas, corresponding to the hopping strength in tight-binding calculations. **Figure 2d** presents measurements of a photonic molecule composed of two adjacent $d = 2$ µm sized mesas of the EnS sample with $h_{\mathrm{EnS}} = 15.3$ nm. For decreasing reduced distances, $v = a/d$, the fundamental *s*-mode and the $p_x$- and $p_y$-modes energetically split into their corresponding bonding and anti-bonding modes. This is well captured by the 2D time-independent Schrödinger equation with a spatial potential $V(x, y)$. The solutions to this eigenvalue problem are plotted as solid lines and reproduce the measurements well.

A significant advantage of the EnS and DnS techniques is that they do not require monocrystalline overgrowth of the top DBR. This permits deeper etching with concurrently smaller feature sizes, which can enable the study of polarization effects in these types of microcavities[38] or the investigation of single-photon emission through the polariton blockade mechanism.[11,12,39]

After establishing both the EnS and DnS techniques, we shift our focus to lattice emulation and demonstrate polariton lasing from a corner defect in a breathing Kagome lattice produced using the EnS method. Importantly, this realization of a corner defect is not equivalent to a higher-order topological insulator, due to its underlying $C_3$ symmetry.[40] The breathing Kagome lattice consists of a three-atomic unit cell with uniform on-site potential and an intra-cell hopping $t_1$. Its inter-cell hopping $t_2$ differs from $t_1$, which is technologically realized as a reduced distance $v_1 = 1.1$ in the unit cell and a $v_2 = 0.85$ in between unit cells. With appropriate termination ($t_1 < t_2$), this results in a staggered hopping along the outer edge that hosts 0D corner modes. An atomic force microscope image of the lattice under investigation is shown in the SI Section 5 and a sketch of the structure in the inset of **Figure 3b**.

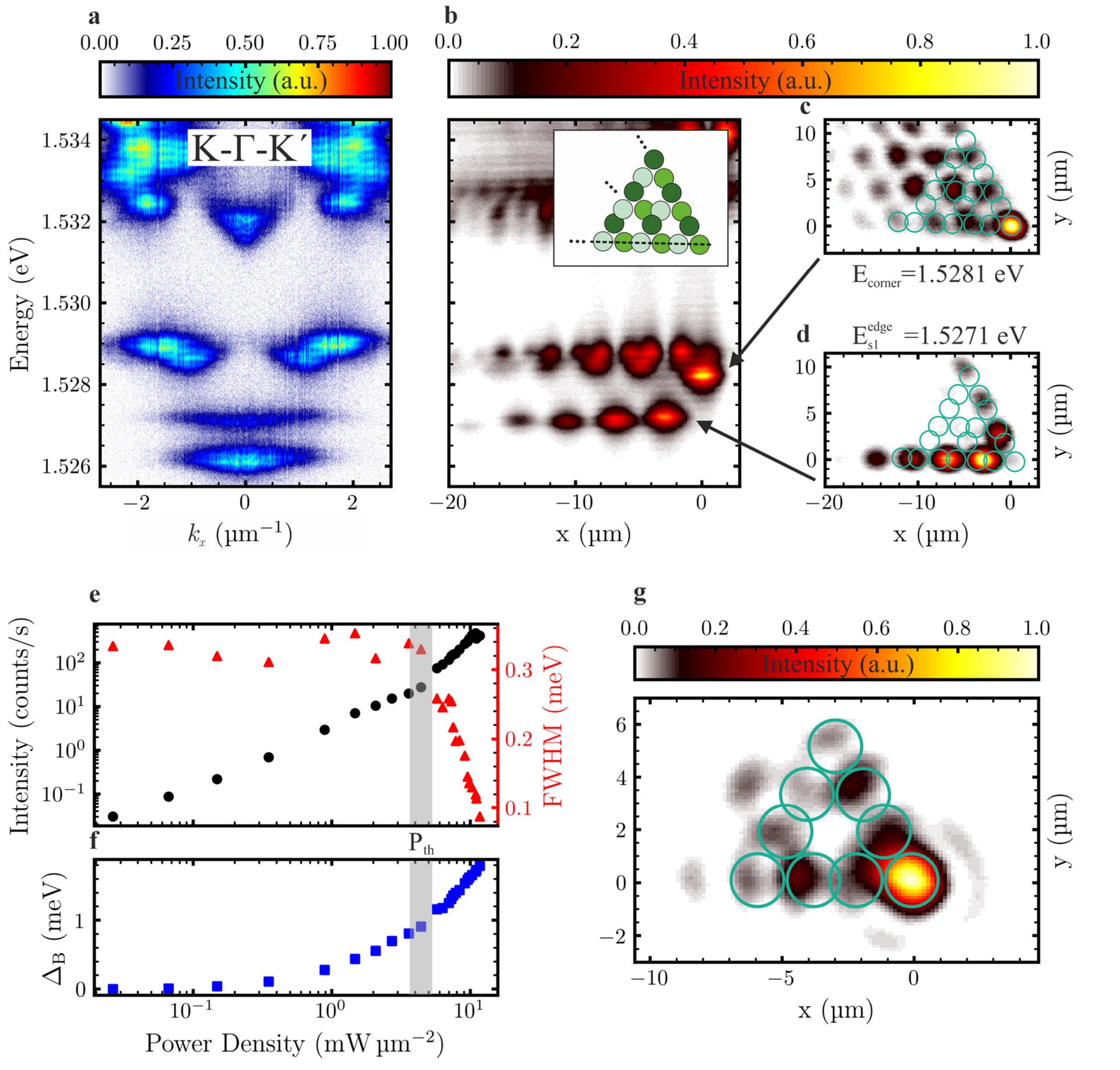

*Figure 3* ***Characterization of the breathing Kagome lattice in the linear and non-linear regime****. Panel* ***a*** *shows the bulk band structure of the lattice along its K-Γ-K' direction, as well as a signature of the 1D edge mode at* $E_{s1}^{edge} = 1.5271$ *eV. The 0D corner mode is spectrally resolved in panel* ***b*** *by a real space cut along the edge of the lattice (see inset). The eigenmode of the 0D corner and 1D edge are spatially resolved at their respective energies in* ***c*** *and* ***d****.* ***e*** *shows features of polariton lasing from the 0D corner, a nonlinear input-output characteristic and a sudden drop in linewidth above the threshold power* $P_{th} \approx 4.4$ *mWμm*$^{-2}$*.* ***f*** *depicts a continuous blueshift* $\Delta_B$ *of the 0D corner mode energy characteristic for polariton lasing and in* ***g*** *the corner mode is spatially resolved above the polariton lasing threshold at* $P = 2.1P_{th}$*. Cyan circles illustrate the underlying lattice geometry.*

The etch depth in this lattice is $h_{\mathrm{EnS}} = 13.4$ nm, the sample is the same as in **Figure 1**. Further details on the breathing Kagome lattice are provided in SI Section 8. Momentum-space spectroscopy reveals the band structure of the polaritonic breathing Kagome lattice along the high symmetry direction K-Γ-K'. For the bulk lattice, we identify the first $s$-band ($E_{s1}^{\mathrm{bulk}} = 1.5262$ eV, bandwidth $\delta E_{s1}^{\mathrm{bulk}} = 110$ μeV), the second $s$-band ($E_{s2}^{\mathrm{bulk}} = 1.5286$ eV, bandwidth $\delta E_{s2}^{\mathrm{bulk}} = 500$ μeV), and a characteristic flatband ($E_{s3}^{\mathrm{bulk}} = 1.5291$ eV), which lies energetically just above the second $s$-band (cf. **Figure 3a**). The termination (cf. Figure 3b) defines an inter-cell distance of $a_{\mathrm{inter}} = 1.7$ μm and an intra-cell distance of $a_{\mathrm{intra}} = 2.2$ μm, giving rise to a 0D corner mode (cf. **Figure 3c**) and a 1D edge mode (cf. **Figure 3d**). The 1D edge channel was clearly resolved in k-space and located at $E_{s1}^{\mathrm{edge}} = 1.5271$ eV, within the bandgap of the bulk. However, due to the weak signal of the 0D corner mode compared to the large bulk contribution its emission is not visible in momentum-space.

To resolve the corner mode, we perform a real-space mode tomography by scanning the lattice corner and extracting an energy-resolved cut along the edge (dashed line in Fig. 3b inset). The higher excitation power caused a blueshift of ΔE = 670 μeV relative to the momentum-space data, which was corrected in **Fig. 3b** for comparison. In **Figure 3b** the bulk modes at $E_{s1}^{\mathrm{bulk}}$ and $E_{s2}^{\mathrm{bulk}}$of the lattice are not visible. Besides the bonding $s$-mode of the edge at $E_{s1}^{\mathrm{edge}}$(cf. **Figure 3d)**, the second $s$-band at $E_{s2}^{\mathrm{edge}} = 1.5288$ eV is also visible. The signal of the 0D corner state is marked in **Figure 3b** and appears at $E_{\mathrm{corner}} = 1.5281$ eV inside the band gap of the bulk mode and edge mode. The corner mode energy can be finetuned by adjusting the lattice coupling parameters.[24] To confirm the presence of a corner-state, the spatial intensity at $E_{\mathrm{corner}}$ is shown in **Figure 3c**, with cyan circles marking mesas and verifying emission from the lattice corner. The signal of the corner-state in **Figure 3b** clearly dominates the real space intensity distribution. Based on its energy position within the bandgap and its spatial localization, this mode is unequivocally identified as the 0D corner mode. These observations are supported by solutions of the Gross-Pitaevskii equation for the lattice potential, shown in SI Section 9.

We now analyze corner-state lasing, as shown in **Figure 3e**. A clear non-linearity in the output intensity and a sudden reduction of linewidth are observed at a threshold power of $P_{\mathrm{th}} \approx 4.4 \mathrm{mW} \mu \mathrm{m}^{-2}$, both hallmark signatures for the onset of coherent emission. A continuous blueshift $\Delta_{\mathrm{B}}$ (cf. **Figure 3f)** across the measured power range confirms that strong coupling is maintained.

To confirm lasing from the corner-state, a real-space mode tomography is performed at a power of $P = 2.1P_{\text{th}}$ (cf. **Figure 3g**). Cyan circles mark the lattice potential, confirming that polariton lasing originates from the corner mode.

## 3. Conclusion

We have introduced two novel fabrication methods for creating uniform potential landscapes in microcavities: the etch-and-oversputter technique, which uses an all-dielectric mirror on top of an etched semiconductor cavity layer, and the deposition-and-oversputter technique, which defines the photonic potential landscape by deposition of $TiO_2$ mesas. Both methods significantly simplify the realization of complex potential landscapes and enable efficient, precise patterning with high spatial fidelity. Using these techniques, we successfully fabricated densely packed Kagome and breathing Kagome lattices and resolved their polaritonic band structures through momentum- and real-space spectroscopy. In the breathing Kagome geometry, we directly observed 1D edge states and 0D corner modes and verified polariton lasing from the corner mode through excitation-power dependence. Our results establish a versatile platform for emulating complex Hamiltonians in a highly controlled environment, including physics of higher-order topology.[9,41] Moreover, the ability to engineer tightly confined resonators with high quality factors represents a significant step toward exploring the polariton blockade mechanism in microcavities with small mode volume.[39]

## Data Availability Statement

The data that support the findings of this study are available from the corresponding authors upon reasonable request.

## Supporting Information

Additional experimental details, materials and methods including the experimental setup, sample fabrication and theory.

## Acknowledgements

We thank Jochen Manara and Thomas Stark from the Center for Applied Energy Research e.V. for ellipsometry characterization measurements.

The Würzburg Team acknowledges financial support by the German Research Foundation (DFG) under Germany's Excellence Strategy–EXC2147 "ct.qmat" (project id 390858490), DFG project KL3124/3-1, as well as DFG project INST 93/1025-1 FUGG. B. P. would like to acknowledge *Quantum Optical Networks based on Exciton-polaritons (Q-ONE)* project funded by HORIZON-EIC-2022_PATHFINDER CHALLENGES EU program under grant agreement No. 101115575. A. W. acknowledges funding support from National Science Center in Poland project No. 2022/47/B/ST3/02411.

## Data availability statement

The data that support the findings of this study are available from the corresponding author upon reasonable request.

## Conflict of Interest

The authors declare no conflict of interest.

## Present Addresses

[#] A.L.: University of Sydney, Camperdown, 2050, Australia.